\documentclass[prb,aps,twocolumn,dcolumn,superscriptaddress,showpacs,amsfonts,amsmath,amssymb,showkeys,floatfix]{revtex4}
\usepackage{bm}
\usepackage{graphicx}
\begin{document}
\title{Equilibrium spin-glass transition of magnetic dipoles with random anisotropy axes on a site diluted lattice}
\date{\today}
\author{Julio F. Fern\'andez}
\affiliation{Instituto de Ciencia de Materiales de Arag\'on, CSIC---Universidad de Zaragoza, 50009-Zaragoza, Spain}
\email[E-mail address: ] {jefe@Unizar.Es}
\author{Juan J. Alonso}
\affiliation{F\'{\i}sica Aplicada I, Universidad de M\'alaga,
29071-M\'alaga, Spain}

\date{\today}

\begin{abstract}
We study partially occupied lattice systems of classical magnetic dipoles which point along randomly oriented axes. Only dipolar interactions are taken into account. The aim of the model is to mimic collective effects in disordered assemblies of magnetic nanoparticles. From tempered Monte Carlo simulations, we obtain the following equilibrium results. The zero temperature entropy approximately vanishes. Below a temperature $T_c$, given by $k_BT_c= (0.95\pm 0.1)x\varepsilon_d $, where $\varepsilon_d$ is a nearest neighbor dipole-dipole interaction energy and $x$ is the site occupancy rate, we find a spin glass phase. In it, (1) the mean value $\langle \mid q\mid\rangle$, where  $q$ is the spin overlap, decreases algebraically with system size $N$ as $N$ increases, and (2) $\delta \mid q\mid  \simeq 0.5 \langle\mid q \mid \rangle\sqrt{T/x}$, independently of $N$, where $\delta \mid q\mid$ is the root mean square deviation of $\mid q \mid$. 
\end{abstract}

\maketitle


\section{introduction}

Magnetic dipole-dipole interactions play a fundamental role in magnetic phenomena. Their long range nature gives rise to magnetic domains in ferromagnets. On the other hand, because of their weak strength, they usually play an insignificant role in determining the Curie temperatures and in critical phenomena of atomic crystals. In assemblies of magnetic nanoparticles (NPs), things can be very different. Because ferromagnetic NPs of up to thousands of Bohr magnetons have a single magnetic domain, \cite{sd} dipole-dipole interactions among such NPs can be very large, and can thus dominate their collective behavior.\cite{dominate} Often, both the orientation of the crystallites of which NPs are made as well as their positions are disordered in the assembly, and therefore behave much as a system of interacting magnetic dipoles with randomly oriented magnetic easy axes. These systems of random axes dipoles (RADs) are clearly frustrated, since two different dipoles give rise to magnetic fields at any given point which are not in general collinear. The sort of time dependent behavior that is expected of spin glasses has been observed in experiments\cite{irrev,2000p,orbach} as well as in simulations\cite{mcaging,ulrich0,ulrich,labarta,bunde} of disordered assemblies of magnetic NPs. Because these systems evolve in time very slowly (exhibiting aging\cite{irrev,mcaging} and other memory\cite{memory} effects), evidence for a \emph {thermodynamic} spin glass phase in them is more difficult to come by. One of us has recently given numerical evidence for the existence of an equilibrium spin glass phase for a \emph{fully} occupied lattice of dipoles with randomly oriented axes.\cite{unos} On the other hand, numerical evidence against such a phase has been given for site \emph{diluted} lattice systems of magnetic dipoles with \emph{parallel} axes.\cite{clare}

One might expect a site diluted systems of RADs to behave as fully occupied ones, and therefore to have equilibrium spin glass phases at low temperatures. For support of this expectation, consider rescaling distances in a dilute system of RADs. Because dipole dipole interactions decay with distance $r$ as $r^{-3}$, letting $r\rightarrow b r$ merely redefines dipole-dipole interaction energies as $\varepsilon_d\rightarrow b^{-3}\varepsilon_d$. This would imply all physical quantities for systems of RAD's in three dimensions (3D) with different values of site concentration $x$ collapse onto the same curve when plotted versus $T/x$, where $T$ is the temperature. This argument holds for randomly located RAD's on a \emph{continuous} space in 3D. It is therefore expected to hold approximately for a site diluted \emph{lattice} if $x\ll 1$, but not necessarily for lattices with higher concentrations of RADs. This is why dipoles with parallel axes can have an antiferromagnetic phase at low temperature on a fully occupied simple cubic lattice,\cite{nos0} but appear to have no condensed phase of any sort on a very dilute lattice.\cite{clare}
Similarly, the existence of a spin glass phase for RADs for $x< 1$ does not follow from its existence for $x=1$.  

Our main aim here is to find, by means of the parallel tempered Monte Carlo (TMC) algorithm,\cite{tempering0,ugrseminar} whether an equilibrium spin glass phase exists in a site diluted system of RADs in 3D. We also aim to establish, if the spin glass phase does exist, whether in the condensed phase (1) there is a single extended state,\cite{extended} as in a ferromagnet or in the droplet model of spin glasses,\cite{droplet} or (2) there are multiple extended states, as in the XY model in 2D\cite{KT} or in the replica symmetry breaking (RSB) theory of spin glasses.\cite{RSB}

The paper is planned as follows. In Sect. \ref{mm} we specify the RAD model and describe how we apply the parallel tempered Monte Carlo (TMC) algorithm.\cite{tempering0,ugrseminar}  We do this for $x=0.35$ and $x=0.5$. Results are given in Sect. \ref{results}. The entropy $S$ follows from our data for the specific heat by numerical integration. It approximately vanishes at zero temperature.\cite{entropy} More precisely, $S<0.01k_B$ at $T=0$. 
We provide numerical evidence for the existence of an equilibrium spin glass phase below a transition temperature $T_c$. The evidence comes from the behavior of the distribution of the spin overlap parameter $q$.\cite{EA,overlap} In analogy to the behavior of the XY model in 2D, $\langle \ q^2\rangle$ seems to vanish in the macroscopic limit in the spin-glass phase, but only as a power of system size. For both $x=0.35$ and $x=0.5$, $k_BT_c= (0.95\pm 0.1)x\varepsilon_d $, where $\varepsilon_d$ is a nearest neighbor (NN) dipole-dipole interaction energy which is defined in Sect. \ref{mm}. (Within errors, the value of $T_c$ for $x=1$\cite{unos} we have previously obtained also satisfies this expression.) For $T\lesssim 0.9T_c$, $\delta \mid q \mid$, that is, the root mean square deviation of $\mid q\mid$, fulfills $ \delta \mid q\mid  \simeq 0.5 \langle\mid q \mid \rangle\sqrt{T/T_c}$, independently of system size. This result is compared with its counterpart for the XY model in 2D.
Results are discussed in Sect. \ref{disc}.

\section{model and method}
\label{mm}

We treat systems of magnetic dipoles on simple cubic lattices. We place a dipole on each lattice site with probability $x$, and leave the site unoccupied with probability $1-x$. Each dipole points along a randomly oriented anisotropy axes. 
With this model we aim to mimic assemblies of NPs in which uniaxial anisotropy energies are an order of magnitude larger than the largest dipole-dipole energy. Then, anisotropy energy barriers are not so large as to freeze spin reorientations near the spin glass temperature, but are sufficiently large to restrict spins to point ``up'' or ``down'' approximately along the easy magnetization axes. We term this the random-axes-dipolar (RAD) model.
The Hamiltonian is given by,
\begin{equation}
{\cal H}=\frac{1}{2}\sum_{ ij}\sum_{\alpha\beta}
T_{ij}^{\alpha\beta}S_i^\alpha S_j^\beta
\end{equation}
where the first sum is over all occupied sites $i$ and $j$ of a simple cubic lattice, $S_i^\alpha$ is the $\alpha$ component of a classical 3-component spin on site $i$, 
\begin{equation}
T_{ij}^{\alpha\beta}=\varepsilon_d
(a/r_{ij})^3(\delta_{\alpha\beta}-3
r_{ij}^\alpha r_{ij}^\beta/r_{ij}^2),
\label{T}
\end{equation} 
$ r_{ij}$ is the distance between $i$ and $j$, $\varepsilon_d$ is an energy, and $a$ a nearest neighbor distance. Each spin points along a randomly chosen direction. These equations can be cast into a form that is manifestly Ising like by letting $\textbf{ \^u}_j$ be (1) a null vector if the $j$ site is unoccupied and (2) a $3-$component unit vector chosen randomly from a spherically uniform distribution if the $j$ site is occupied, and defining a pseudospin $\sigma_j=\pm 1$ for each site, such that $\textbf{ S}_j=\textbf{ \^u}_j\sigma_j$. 
We can then write,
\begin{equation}
{\cal H}=-\frac{1}{2}\sum_{ ij}J_{ij}\sigma_i\sigma_j,
\label{Hr}
\end{equation}
where
$J_{ij}=-\sum_{\alpha ,\beta}T_{ij}^{\alpha\beta}u^{\alpha}_i
u^{\beta}_j$. Thus, the RAD model is an Ising model whose bonds $J_{ij}$ are determined by the dipole-dipole terms $T_{ij}^{\alpha\beta}$ and the set of 3-component randomly oriented unit vectors $\{\textbf{\^u}_j\}$.

From here on, unless we state otherwise, we let  $k_B=1$, where  $k_B$ is Boltzmann's constant, and give all temperatures in terms of $\varepsilon_d$.  

We use periodic boundary conditions. Details and justification are given in Refs. [\onlinecite{unos,nos0,ugrseminar,08}].
\begin{table}
\caption{Simulation parameters. $x$ is the probability that any given site is occupied by a magnetic dipole; $N$ is the mean number of magnetic dipoles in the system; $\Delta T$ is the temperature step in the TMC runs; $T_m$ is the highest temperature of all systems; $N_s$ is the number of disordered system pairs; MCS is the number of tempered Monte Carlo sweeps.}
\begin{ruledtabular}
\begin{tabular}{|r|r|r|r|r|r|r|r|r|}
x & 0.35 & 0.35 & 0.35 & 0.35 & 0.5 & 0.5  & 0.5 & 0.5\\ 
$N$ & 22.4 & 75.6 & 179.2 & 604.8  & 32 & 108 & 256 & 864 \\
$\Delta T$ &  0.05 & 0.05 & 0.05 &  0.05 & 0.05 & 0.05 & 0.05 & 0.05 \\
$T_m$ &  1.0 & 1.0 & 1.0 &  1.0 & 1.3 &1.3 & 1.3 & 1.3 \\
$N_s$ & 10000 & 3000 & 1250 & 900  & 5000 & 3000 & 1500 & 150,600 \\
MCS & $50 000$&$10^5$&$5\times 10^6$& $10^6$ &$50 000$&$50 000$&$10^6$&$ 10^6, 10^5$\\
\end{tabular}
\end{ruledtabular}
\end{table}

In order to arrive at equilibrium results, we make use of the parallel tempered Monte Carlo (TMC) algorithm.\cite{tempering0}
This enables one to circumvent large energy barriers that can trap a system's state. We apply the TMC algorithm as follows. We run (in parallel) several identical systems at different temperatures: a system at temperature $T_0$, a second one at $T_0+\Delta T$, and so on, at equally spaced temperatures, up to $T_m$ 
We choose $T_m$ to be at least twice as large as what we expect to be the transition temperature between the paramagnetic and spin glass phases. We let each system evolve under the Metropolis MC algorithm for 10 MC sweeps before pairs of systems are given a chance to exchange their states. More specifically, pairs of systems at temperatures ($T_0$ and $T_0+\Delta T$), ($T_0+2\Delta T$ and $T_0+3\Delta T$), and so on, are given a chance to exchange states every $10$ MC sweeps, and pairs of systems at temperatures ($T_0+\Delta T$ and $T_0+2\Delta T$), ($T_0+3\Delta T$ and $T_0+4\Delta T$), and so on, are given a chance to exchange states at times in between. Under TMC rules, these chances are as follows. A system at temperature $T$ that is in state $2$, and another one at $T+\Delta T$ in state $1$, exchange states with probability $P=1$ if $E_1<E_2$ and $\Delta T>0$, but $P=\exp (-\Delta \beta \delta E)$, where $\Delta \beta=1/T-1/(T+\Delta T)$ and $\delta E=E_1-E_2$, if $E_1>E_2$ and $\Delta T>0$.\cite{ugrseminar} 

\begin{figure}[!ht]
\includegraphics*[width=80mm]{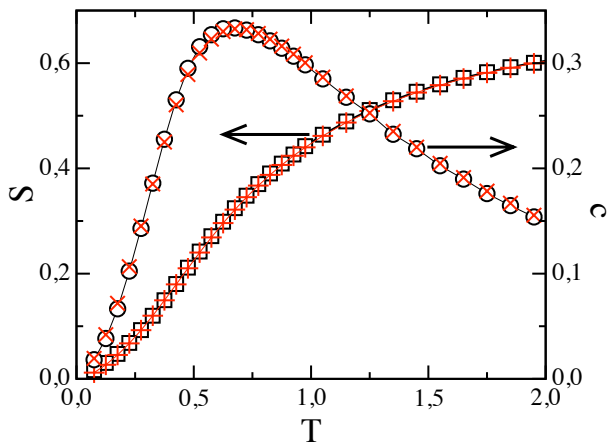}
\caption{(a) (Color online) Plots of  $c$ and of $S$ vs $T$, for $x=0.35$. Data points (black) $\circ$ (for $N=605$) and (red) $\times$ (for $N=179$) are for the specific heat, while (black) $\square$ (for $N=605$) and (red) $+$ (for $N=179$) are for the entropy.
Errors, $\delta S$, for $S$ increase as $T$ decreases, from $\delta S\approx 0.002$ for $T=4$ to $\delta S\approx 0.004$ for $T<0.1$.}
\label{CyS}
\end{figure}

A sufficiently small value of $\Delta T$ must be chosen in order to keep $\exp (-\Delta \beta \delta E)$ from becoming too small. This will often be fulfilled if $\Delta \beta\Delta E\lesssim 1$, where $\Delta E$ is the mean energy difference between two systems at temperatures $T$ and $T+\Delta T$. The required condition, $\Delta T \lesssim T/\sqrt{Nc}$, where $N$ is the number of dipoles in the system, follows for $\Delta T$  if we replace $\Delta E$ by $cN\Delta T$ (we thus define $c$ as the specific heat per spin). From plots of the specific heat vs $T$, such as the one shown in Fig. \ref{CyS} for systems of $179$ and of $605$ magnetic dipoles on lattices with $0.35$ of their sites occupied, one can get upper bounds for $\Delta T$. (How the data points shown in Fig. \ref{CyS} were obtained is explained in Sect. \ref{results}.) Values of $\Delta T$ as well as of other parameters for all TMC runs are given in table I.

In order to probe for spin glass behavior, we define, as is usual, the spin overlap parameter.\cite{EA,overlap} First, let 
\begin{equation} 
\phi_j=\sigma^{(1)}_j\sigma^{(2)}_j,
\label{phi}
\end{equation}
where $\sigma^{(1)}_j$ and $\sigma^{(2)}_j$ are the pseudospins [defined above Eq. (\ref{Hr})] on site $j$ of identical twin replicas 1 and 2 of the system.
Clearly, 
\begin{equation} 
q= N^{-1}\sum_j\phi_j
\label{qtilde}
\end{equation}
is a measure of the spin configuration overlap between replicas 1 and 2. Thus,
$\mid q\mid=1$ if either $\sigma_j^{(1)}= \sigma_j^{(2)}$ for all $j$ or $\sigma_j^{(1)}=- \sigma_j^{(2)}$ for all $j$. 
We also define the moments of $q$, 
$q_k= \langle \mid q \mid ^k  \rangle$,
for $k=1$, $2$ and $4$, where $\langle \ldots\rangle$ stands for an average which we next specify. 

Suppose our TMC runs last for a time $t$, and assume that, within the temperature range we are interested in, equilibration takes place in a time less than $t/2$. During the TMC runs, we write, at equally spaced time intervals within the $(t/2,t)$ range, the spin configuration for each temperature. We later draw from these written configurations the numbers for $\sigma^{(1)}_j$ that go into Eq. (\ref{phi}). This procedure is repeated for the same system (that is, for a system with the same anisotropy axes), but starting from different initial conditions. From this second set of configurations, we draw the numbers for $\sigma^{(2)}_j$ that go into Eq. (\ref{phi}). A value of $q$ for each time $t$ and temperature $T$ is thus obtained from Eq. (\ref{qtilde}). We first average all quantities of interest over $t$, and finally repeat these pairs of TMC runs a number $N_s$ of times for different realizations of axes orientations, from which the average values we are reporting  are obtained. Values we have used in our TMC runs for $N_s$ and for $t$ (labeled MCS therein) are given in table I. 

Finally, in order to check that equilibration actually takes place as assumed, we also calculate the spin overlap $\tilde {q}$, not between identical twin replicas, but between spin configurations at two different times $t_0$ and $t_1$ of the same TMC run. We do this for several values of $t_0$ and $t_1$ in the neighborhoods of $t/2$ and $t$ respectively, and average over different random axes realizations, in order to obtain $\tilde {q}_1$. For $t$ sufficiently large, $\tilde {q}_1=q_1$. We find this is fulfilled for all parameter values given in table I, 
 except for (1) $x=0.35$, $N=604.8$ and $T\lesssim 0.2$, and (2) for $x=0.5$, $N=864$ and $T\lesssim 0.35$.

\begin{figure}[!ht]
\includegraphics*[width=80mm]{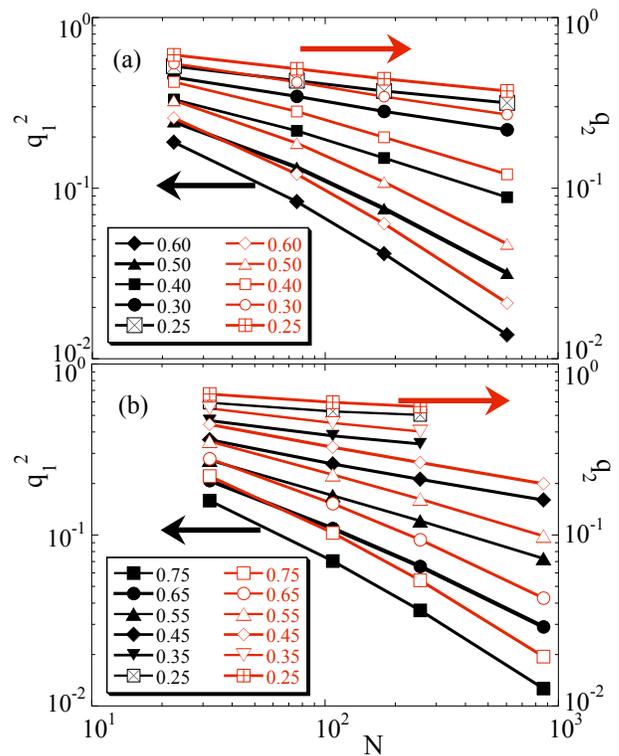}
\caption{(Color online) (a) Log-log plots of $q_1$ and of $q_2$ vs $N$ for $x=0.35$ and the shown values of $T$. Lines are guides to the eye. (b) Same as in (a) but for $x=0.5$. Some data points for small $T$ and large $N$ are missing, because our TMC runs did not reach equilibration for them.}
\label{q2q1}
\end{figure}

\section{results}
\label{results}

Plots of the specific heat and of the entropy vs $T$ for the RAD model are shown in Fig. \ref{CyS} for $x=0.35$. The data for the specific heat follows from numerical derivatives of the energy with respect to $T$. We obtain $S$ from $S(T)=\ln 2+\int_\infty^T c(T^\prime )/T^\prime dT^\prime$. Our data covers the temperature range (not all of it shown in Fig.  \ref{CyS}) $0.05< T<4$. For a numerical integration, we must extrapolate our data for $c(T)$ beyond $T=4$. To this end, we use the leading term of an energy expansion in powers of $1/T$, which gives $c(T) \rightarrow A/T^2$. A fit of the value of $A$ to our data for $c(T)$ at $T=4$ leads to a $\Delta S\simeq 0.027$ contribution to the entropy from the $T>4$ range. We thus obtain $S(T)$. It is exhibited in Fig. \ref{CyS}. 
Errors for $S(T)$ come in two approximately equal pieces: (1) an error of roughly $0.002$ from the $4<T<\infty$ range, and (2) an error of roughly $0.002$, from 
errors in the data for the specific, which enter the integral $\int_\infty^T c(T^\prime )/T^\prime dT^\prime$. We finally obtain $S=0.015\pm 0.004$ for $T=0.05$, and extrapolations below $T=0.05$ yield $S<0.01$ at $T=0$. Similarly, we find $S<0.01$ at $T=0$ for $x=0.50$.

\begin{figure}[!ht]
\includegraphics*[width=80mm]{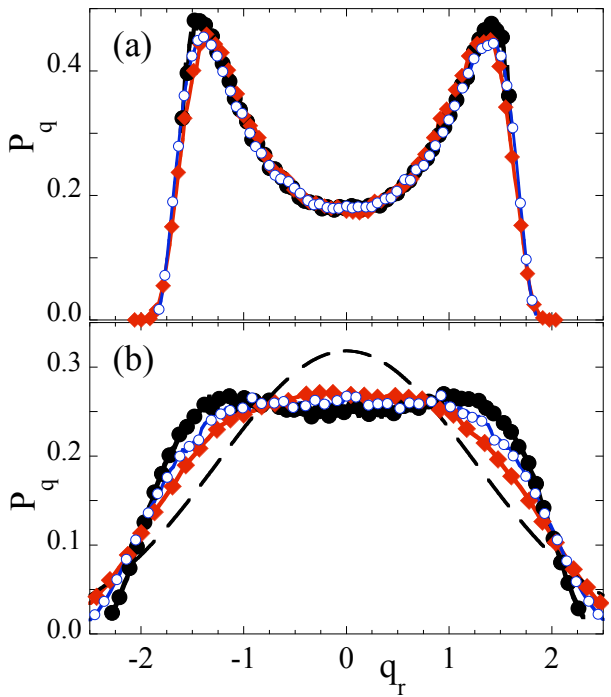}
\caption{(Color online) (a) Plots of the probability distribution $P_q$ vs $q_r$, for $x=0.35$ and $T=0.3$. $\bullet$, $\circ$, and $\blacklozenge$, are for $N=76, 179$ and $605$, respectively. Lines are guides to the eye. For clarity, only 30\% of the data points are shown. (b) Same as in (a) but for $T=0.45$. For comparison, a dashed line is shown for $(1/\pi )\exp(-q_r^2/\pi)$, which ensues for a macroscopic paramagnetic phase.}
\label{pqx35}
\end{figure}  
\begin{figure}[!ht]
\includegraphics*[width=80mm]{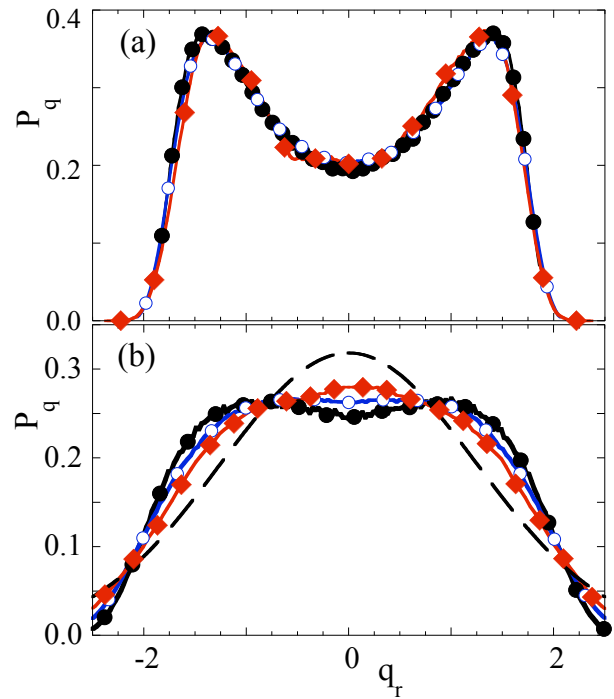}
\caption{(Color online) (a) Plots of the probability $P_q$ vs $q_r$, for $x=0.50$ and $T=0.45$. $\bullet$, $\circ$, and $\blacklozenge$, are for $N=108, 256$ and $864$, respectively. Lines are guides to the eye. For clarity, only 30\% of the data points are shown. (b) Same as in (a) but for $T=0.60$. For comparison, a dashed line is shown for $(1/\pi )\exp(-q_r^2/\pi)$, which ensues for a macroscopic paramagnetic phase.}
\label{pqx50}
\end{figure}

Our results for $q_1$ and $q_2$ follow.
Note we use an absolute value in the definition of $q_1$. Recall that $Nq_2=\it {O}(1)$ in the paramagnetic phase, and diverges at the transition temperature $T_c$. Above the lower critical dimension $d_c$, $q_2=\it {O}(1)$ in the droplet model of the spin glass phase.\cite{droplet} 
Our data shows that $q_2$ decreases as $N$ increases, for all nonzero temperatures. Log-log plots of $q_2$ vs $N$ are exhibited in Figs. \ref{q2q1}a and \ref{q2q1}b for various values of $T$. The behavior of $q_2$ for $T\lesssim 0.3$ and $T\lesssim 0.45$ for $x=0.35$ and $x=0.5$, respectively, is consistent with  $q_2\sim N^{-y}$, where $y$ is some positive parameter that depends on $T$. This behavior is reminiscent of the XY model in 2D.\cite{KT} 
For higher values of $T$, $q_2$ vs $N$ clearly curves downwards, in accordance with a faster than algebraic in $N$ decay, as one expects for the paramagnetic phase. Log-log plots of $q_1^2$ are also shown in Fig. \ref{q2q1}a and \ref{q2q1}b for various values of $T$. Note that $q_2/q_1^2>1$ for all $N$ and nonzero $T$, which implies a nonvanishing uncertainty in $\mid q_r\mid$, where $q_r=q/q_1$, for $T>0$.

We next report results for the probability distribution $P_q(q_r)$. Unless stated otherwise, all data given below are for a normalized $P_q(q_r)$, that is, $\int P_q(q_r)dq_r =1$.
Recall that in the paramagnetic phase, because spin-spin correlation lengths are finite, the central limit theorem implies $q_r$ is normally distributed for macroscopic systems. 
Plots of $P_q$ vs $q_r$ are shown in Fig. \ref{pqx50}a (for $T=0.45$) and Fig. \ref{pqx50}b (for $T=0.60$), both for $x=0.50$. The distribution of $q$ appears to be size independent for $T=0.45$. 
For $T=0.60$, on the other hand, $P_q(q_r)$ drifts with system size. The drift seems consistent with  $P_q(q_r)\rightarrow (1/\pi )\exp (-q_r^2/\pi )$ as $N\rightarrow \infty$, which would be in accordance with a paramagnetic phase.  
Similar remarks apply to the plots exhibited in Fig. \ref{pqx35}a (for $T=0.30$) and Fig. \ref{pqx35}b (for $T=0.45$), both for $x=0.35$. Clearly, $0.30<T_c<0.45$ for $x=0.35$, and $0.45<T_c<0.60$ for $x=0.50$. For $T<T_c$, plots of $P_q$ vs $q_r$ (not shown) become sharply peaked only as $T\rightarrow 0$. 
Now, if it turns out that $d_c=3$ for systems of RADs, then $P_q(q_r)$ ought to exhibit critical behavior for all $T\leq T_c$, more specifically, $P_q(q_r)$ ought to be size independent.
This sort of behavior is best summarized by the mean square deviation of $\mid q_r \mid$, $\Delta_q^2\equiv \langle q_r^2\rangle - \langle \mid q_r\mid \rangle^2$.

\begin{figure}[!ht]
\includegraphics*[width=80mm]{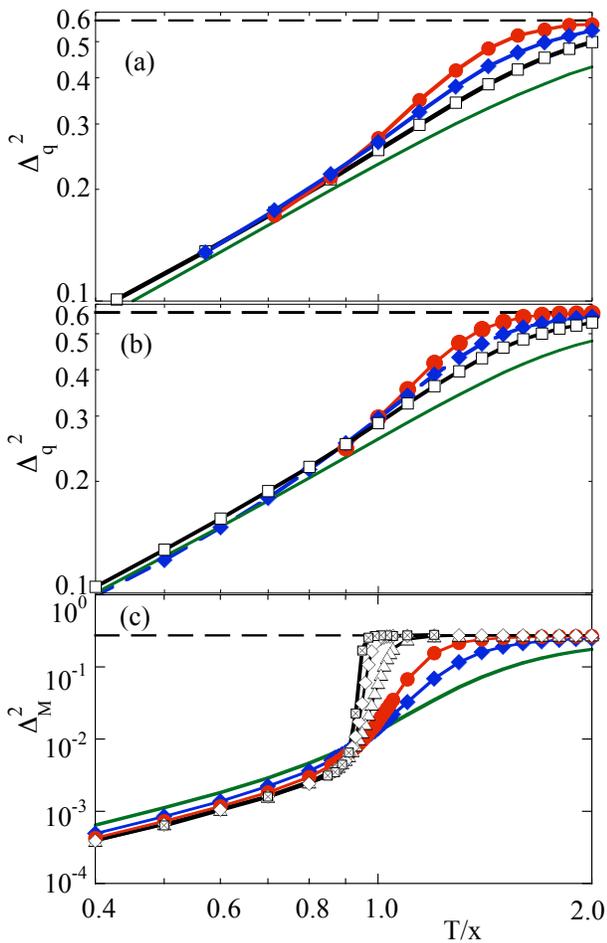}
\caption{(Color online) (a) Plots of $\Delta^2_q$ vs $T$ for $x=0.35$. (red) $\bullet$, (blue) $\blacklozenge$, and (black) $\square$ are for $N =604.8, 179.2$, and $75.6$, respectively. The continuous (green) line is for $N = 22.4$. The dashed line is for macroscopic paramagnet. (b) Same as in (a) but for $x=0.50$. $\bullet$ (red), $\blacklozenge$ (blue), and $\square$ (black) are for $N = 864, 256$, and $108$, respectively. The continuous (green) line is for $N = 32$, and the dashed line is for $\pi /2-1$. (c) Same as in (a), but for the mean square relative deviation $\Delta _M^2$ of $\mid M\mid$, where $M$ is the magnetic moment of XY systems of $L\times L$ spins in 2D vs $T/x$, where $x=1$, for $L=4$ (green line), $L=8$ (blue $\blacklozenge$), $L=16$ (red $\bullet$), $L=64$ (black $\triangle$),  $L=256$ (black $\lozenge$), and L=1024 (black $\boxtimes$).}
\label{Dq}
\end{figure} 

Plots of $\Delta_q^2$ vs $T/x$ are shown in Fig. \ref{Dq}a (for $x=0.35$) and in Fig. \ref{Dq}b for $x=0.50$ for various values of $N$. Clearly, curves in Figs. \ref{Dq}a
and \ref{Dq}b differ only slightly, as expected from the argument given in Sect. I, that variations of $T$ and $x$ have an effect only through $T/x$ if $x\ll 1$. The data points in both figures suggest $\Delta_q^2\rightarrow \pi /2-1$ as $N\rightarrow\infty$, for $T/x\gtrsim 1$. This is expected for a macroscopic paramagnet, since $P_q(q_r)$ is a normal distribution then.
For $T/x\lesssim 0.8$, 
\begin{equation}
\Delta_q \simeq 0.5  \sqrt{T/x}
\label{dqsq}
\end{equation}
provides the best fit to the data points shown in Figs.  \ref{Dq}a and  \ref{Dq}b. This is in contrast with the behavior,  $\Delta_q^2\rightarrow 0$ as $N\rightarrow\infty$, that one expects for the ordered phase of ferromagnets such as the Ising model in two or higher dimensions. For a more relevant comparison, we show in Fig. \ref{Dq}c plots of the mean square relative deviation $\Delta _M^2$ of $\mid M\mid$, where $M$ is the magnetic moment, of the XY model in $2D$. Notice there is no counterpart in Figs. \ref{Dq}a and \ref{Dq}b for the data points shown for rather large systems in Fig. \ref{Dq}c. We note that $\Delta_M^2$ seems to go to $ 4\times 10^{-3}T^{2.2}$ as $N\rightarrow \infty$, for $T<T_c$ in the XY model in $2D$.

\begin{figure}[!ht]
\includegraphics*[width=80mm]{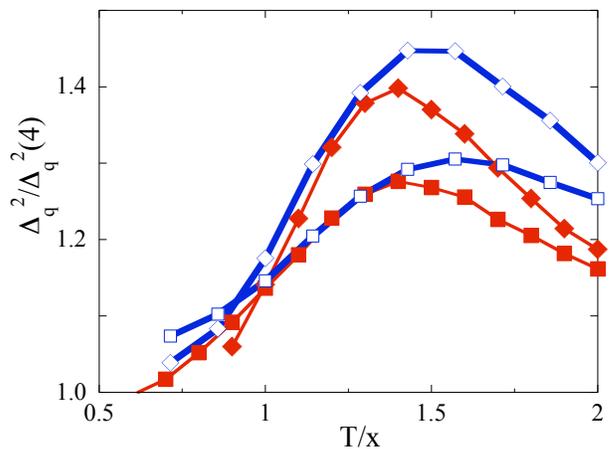}
\caption{(Color online) Plots of $\Delta^2_q/\Delta^2_q(L=4)$ vs $T$, where $\Delta^2_q(L=4)$ is $\Delta^2_q$ for $L=4$.
All ($\blacklozenge$) are for $x=0.35$ and $L=12$, and all ($\blacksquare$) are for $x=0.35$ and $L=8$. All ($\Diamond$) are for $x=0.50$ and $L=12$, and all ($\square$) are for $x=0.50$ and $L=8$. Lines are guides to the eye.}
\label{ratio}
\end{figure}

Now, it is hard to see in Figs. \ref{Dq}a and \ref{Dq}b where curves merge or cross, near $T/x=0.9$. In order to enhance the differences between $\Delta_q^2$ curves for different values of $L$, we plot $\Delta_q^2/\Delta_q^2(L=4)$ vs $T/x$. These plots are shown in Fig. \ref{ratio} for $x=0.35$ and $x=0.50$. The weak dependence on $x$ is remarkable. Notice that whereas curves for $L=12$ and $L=8$ cross near $T/x=0.95$ for $x=0.35$ and for $x=0.50$, the same curves cross the $\Delta_q^2/\Delta_q^2(L=4)=1$ horizontal line only, if at all, for much smaller values of $T/x$. It thus appears that curves for $\Delta_q^2$ increasingly larger system sizes merge or cross at increasingly larger values of $T/x$.

\begin{figure}[!ht]
\includegraphics*[width=80mm]{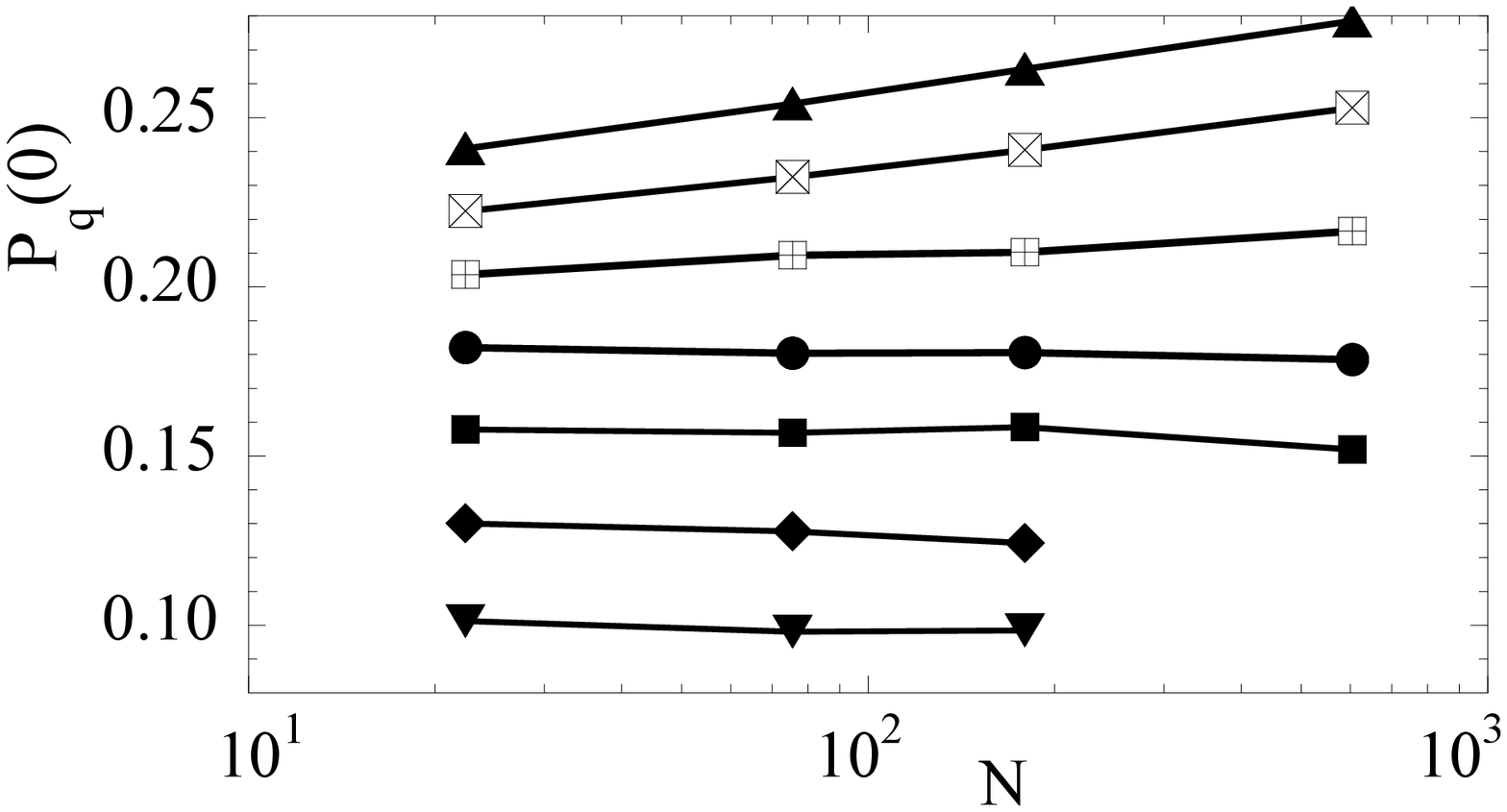}
\caption{ Plots of $P_q(q=0)$ vs $N$, for $x=0.35$, for $T/x=1.286$ ($\blacktriangle$), $T/x=1.143$ ( $\boxtimes$), $T/x=1.0$ ( $\boxplus$), $T/x=0.857$ ( $\bullet$),  $T/x=0.714$ ( $\blacksquare$), $T/x=0.57$ ( $\blacklozenge$), and $T/x=0.429$, ($\blacktriangledown$). }
\label{P0}
\end{figure}

We show plots of $P_q(0)$ vs $N$ for $x=0.35$ and various temperatures in Fig. \ref{P0}. In the paramagnetic phase, we expect $P_q(0)\rightarrow 1/\pi $ as $N \rightarrow\infty$, as follows from a normal distribution of $q_r$ about $q_r=0$. Now, for $T/x\geq 1.0$, we see that $P_q(0)<1/\pi$, but $P_q(0)$ increases as $N$ increases. On the other hand, for $T/x\leq  0.857$, $P_q(0)$ remains, within errors, constant. This is again in contrast with the behavior, $P_q(0)\rightarrow 0$ as $N\rightarrow\infty$, that one expects for the ordered phase of ferromagnets such as the Ising model in two or higher dimensions, but is in accordance with $d_c\simeq 3$.

Finally, taking into account all our observations above, we arrive at our best estimate for $T_c$: $T_c/x=0.95\pm0.1$.

\section{discussion}
\label{disc}

By tempered Monte Carlo calculations, we have studied systems of RADs in simple cubic lattices in which each site is occupied by a magnetic dipole with probability $x$. Systems sizes, Monte Carlo run lengths and other details about the calculations can be found in Table I. 
The entropy $S$ as a function of temperature, which follows from our data for the specific heat, is shown in Fig. \ref{CyS} for $x=0.35$. $S$ approximately vanishes at zero temperature. More precisely, $S<0.01k_B$ at $T=0$ for both $x=0.35$ and $x=0.50$.\cite{entropy} Note also that a vanishing zero temperature entropy implies that trapping above the lowest energy states as $T\rightarrow 0$ occurs only very rarely in our TMC runs.

We have obtained the spin overlap parameter, defined in Eq. (\ref{qtilde}). More specifically, we have obtained the mean value of $\mid q\mid$, which we term $q_1$, and the mean square value of $q$, that is $q_2$. The plots in Fig. \ref{q2q1}a (for $x=0.35$) and in Fig. \ref{q2q1}b (for $x=0.50$) suggest both $q_1$ and $q_2$ vanish for all nonzero $T$ as $N\rightarrow\infty$. As can be gathered from those plots, both $q_1^2\sim N^{-y}$ and $q_2\sim N^{-y}$, where $y$ is some positive parameter that depends on $T$, for $k_BT/x\lesssim 0.9\varepsilon_d $.
This is reminiscent of the XY model in 2D and suggests the lower critical dimension $d_c$ of the RAD model is at or near 3D. 

We have also studied the probability distribution $P_q(q_r)$, where $q_r=q/q_1$. Within errors, $P_q(q_r)$ is independent of system size at $T=T_c$, as is illustrated in Figs. \ref{pqx35}a and  \ref{pqx50}a for $k_BT/x\simeq 0.9\varepsilon_d $. We have also found (not shown) $P_q(q_r)$ to be independent of system size at lower temperatures. The results shown in Figs. \ref{Dq}a and \ref{Dq}b and in Eq. (\ref{dqsq}), for the mean square deviation of $\mid q_r\mid$, $\Delta_q^2$, are consistent with a $P_q(q_r)$ that is independent of system size for all  $k_BT/x\lesssim 0.9\varepsilon_d $. 
Again, this is as in the XY model in 2D (see Fig. \ref{Dq}c) and suggests $d_c\simeq 3$ for the RAD model. 

There is a fairly large quantitative difference between the otherwise analogous behavior of the condensed phases of the XY model in 2D and the RAD model in 3D. 
From Eq. (\ref{Dq}) for the RAD system, and from $\Delta_M^2\simeq 4\times 10^{-3}T^{2.2}$, which we drew from the plots shown in Fig. \ref{Dq}c  for the XY model, $\Delta_q/\Delta_M\approx 10t^{-0.6}$, where $t=T/T_c$, follows. Not surprisingly then, $P_M/M$ at $M\sim 0$ for the XY model\cite{unpublished} is orders of magnitude smaller than $P_q(0)$ for the RAD model, when they are both at some $T/ T_c$ well below a value of $ 1$.
Replica symmetry breaking would account this large quantitative difference between the XY and the RAD models at or near their lower critical dimensionality.

\acknowledgments
The Instituto Carlos I  (at Universidad de Granada)
and the BIFI Institute (at Universidad de Zaragoza) have generously allowed us to run on many of their computer cluster nodes. We received financial support, through Grant No. FIS2006-00708, from the Ministerio de Ciencia e Innovaci\'on of Spain.

\end{document}